\documentclass[twocolumn,showpacs,preprintnumbers,amsmath,amssymb]{revtex4}

\usepackage{graphicx}
\usepackage{dcolumn}
\usepackage{bm}

\begin{document}

\preprint{}

\title{Quantum Noise in Differential-type Gravitational-Wave Interferometer\\and Signal Recycling}

\author{Atsushi Nishizawa}
\email{atsushi.nishizawa@nao.ac.jp}
\affiliation{Graduate School of Human and Environmental Studies, Kyoto University, Kyoto 606-8501, Japan}
\author{Seiji Kawamura}
\email{seiji.kawamura@nao.ac.jp}
\affiliation{TAMA Project, National Astronomical Observatory of Japan, 2-21-1 Osawa, Mitaka, Tokyo 181-8588, Japan}
\author{Masa-aki Sakagami}
\email{sakagami@grav.mbox.media.kyoto-u.ac.jp}
\affiliation{Graduate School of Human and Environmental Studies, Kyoto University, Kyoto 606-8501, Japan}

\date{\today}

\begin{abstract}
There exists the standard quantum limit (SQL), derived from Heisenberg's uncertainty relation, in the sensitivity of laser interferometer gravitational-wave detectors. However, in the context of a full quantum-mechanical approach, SQL can be overcome using the correlation of shot noise and radiation-pressure noise. So far, signal recycling, which is one of the methods to overcome SQL, is considered only in a recombined-type interferometer such as Advanced-LIGO, LCGT, and GEO600. In this paper, we investigated quantum noise and the possibility of signal recycling in a differential-type interferometer. As a result, we found that signal recycling is possible and creates at most three dips in the sensitivity curve of the detector. Then, taking advantage of the third additional dip and comparing the sensitivity of a differential-type interferometer with that of a next-generation Japanese GW interferometer, LCGT, we found that SNR of inspiral binary is improved by a factor of $\approx$1.43 for neutron star binary, $\approx$2.28 for 50$M_{\odot}$ black hole binary, and $\approx$2.94 for 100$M_{\odot}$ black hole binary. We also found that power recycling to increase laser power is possible in our signal-recycling configuration of a detector.
\end{abstract}

\pacs{Valid PACS appear here}
\maketitle

\section{Intoroduction}
The first generation of kilometer-scale, ground-based laser interferometer gravitational-wave (GW) detectors, located in the United States (LIGO), Europe (VIRGO and GEO 600), and Japan (TAMA 300), has begun its search for gravitational wave radiation and has yielded scientific results. The development of interferometers of the next-generation, such as Advanced-LIGO (in U.S.) \cite{bib:10} and LCGT (in Japan) \cite{bib:11}, is underway.\\
\hspace{3mm}In the first-generation interferometers, we can ignore the contribution of radiation-pressure noise because the laser power is low enough. In the next-generation interferometers, laser power is so high that radiation-pressure noise should be treated correctly in a fully quantum-mechanical way, in which the radiation-pressure noise could have the correlation with shot noise \cite{bib:1}. These two noises have different dependences on laser power $I_0$. The spectral density of radiation-pressure noise is proportional to $I_0$ and that of shot noise is inverse-proportional to $I_0$. Thus, there exists an optimal laser power to reach maximum sensitivity at a certain frequency. This maximum reachable sensitivity is called the standard quantum limit (SQL).\\
\hspace{3mm}SQL is alternatively derived from Heisenberg's uncertainty relation on a free mass \cite{bib:9}. What we want to measure is the position of a free mass. However, too accurate measurement on the test mass will greatly perturb the velocity of the test mass and cause the large uncertainty of the position at the next measurement. Therefore, it is indicated that there exists the optimal accuracy and maximum reachable sensitivity of a measurement. This is the more fundamental explanation for the SQL in an interferometer.\\
\hspace{3mm}Nonetheless, it is possible to circumvent SQL by using signal recycling, which is one of the methods circumventing SQL and uses one extra mirror, called a Signal Recycling (SR) mirror\footnote{The interferometer configuration is called signal recycling, detuned signal recycling or resonant sideband extraction, depending on the microscopic position of the extra mirror at the dark port. In this paper, however, we stick to the term "signal recycling" for convenience.}. This additional mirror can reshape the noise curve and make dips in it \cite{bib:12,bib:13}. Recently, quantum noise has been calculated in the case of Advanced-LIGO, in which SR mirror is put at the dark port of the interferometer \cite{bib:2,bib:3}. SR mirror creates dynamical correlations between shot noise and radiation-pressure noise and this makes it possible to circumvent SQL. The signal recycling is planned to apply to the next-generation interferometers such as Advanced-LIGO and LCGT \cite{bib:10,bib:11}. These interferometers interfere two lights returning from two arms and detect a differential signal. This detection method is called recombined-type.\\
\hspace{3mm}On the other hand, there exists another method called differential-type, which detects signals for each arm independently and combines (differentiates) them after the detection. In this configuration, however, one cannot increase laser power using a power-recycling mirror. This seems to be a fatal defect for a differential-type interferometer when it is applied to ground-based interferometers, because more laser power is needed to decrease shot noise. However, in our signal-recycling configuration described in this paper, the power recycling is possible with two SR mirrors located at the output. A differential-type also has the advantage that one can take various combinations of signals in the situation of many detectors. Thus, a differential-type interferometer could become a new design for future GW detectors.\\
\hspace{3mm}The aim of this paper is to investigate quantum noise and achievable sensitivity in differential-type interferometers with signal recycling. In Sec. II, we will review previous work on quantum formalism \cite{bib:1} and signal recycling in a recombined-type interferometer \cite{bib:2,bib:3}. In Sec. III, we will explain the configuration of a differential-type interferometer and derive the spectral density for the cases without and with SR mirrors. The details of the calculation are given in the Appendix. Next, in Sec. IV, based on the results obtained in the previous section, we will decompose full spectral density into three noise parts and describe the physical interpretation of their interesting features. And then, in Sec. V, comparing a differential-type interferometer with a recombined-type one, we will calculate the SNR of inspiral binary stars and show the advantages of our configuration. Finally, Sec. VI is devoted to a summary of this paper and discussions.\\            

\section{Quantum noise in Recombined-type interferometer}        
\subsection{Quantum Formalism}
Recently, full-quantum treatment of quantum noise has been formulated by Kimble et al. (hereafter "KLMTV") \cite{bib:1}. In this subsection, we will review it briefly. Quantum noise is caused by vacuum field $\mathbf{a}$ entering an interferometer from the dark port \cite{bib:15,bib:16}. The $\mathbf{a}$ field is shot noise itself and also produces radiation-pressure noise, coupled with the carrier light in the Fabry-Perot (FP) cavity. $\mathbf{b}$ is the output field, which includes shot noise, radiation-pressure noise and GW signal. \\
\hspace{3mm}An input electric field is written using two-photon mode \cite{bib:5,bib:6}. In this picture, the electric field is written with quadrature modes as,
\begin{eqnarray}
E_{\rm in} &=& \sqrt{\frac{4\pi\hbar\omega_o}{{\cal A}c}} \left[ 
\cos(\omega_o t)
\int_0^\infty \left(a_1 e^{-\i\Omega t} + a_1^{\dag} e^{+i\Omega t}
\right) {d\Omega\over2\pi}\right. \nonumber\\
&&\quad\quad + \left. \sin(\omega_o t) 
\int_0^\infty \left(a_2 e^{-\i\Omega t} + a_2^{\dag} e^{+i\Omega t}
\right) {d\Omega\over2\pi} \right]\;, \nonumber \\
&&
\label{eqn:26}
\end{eqnarray}
where $c$ is the speed of light, $\hbar$ is the reduced Plank constant, $\omega_0$ is the angular frequency of carrier light and ${\cal A}$ is the effective cross-section of a beam. $a_1$ and $a_2$ are fields amplitudes for each quadrature mode. They are defined using the sideband's annihilation operator as,
\begin{equation}
a_1 \equiv {a_+ + a_-^{\dag} \over \sqrt2}\;, \quad 
a_2 \equiv {a_+ - a_-^{\dag} \over \sqrt2 i}\;.
\label{eqn:6}
\end{equation}
where $a_+$ and $a_-$ are annihilation operators for sidebands $\omega_0 \pm \Omega$ and satisfy the ordinary commutation relation
\begin{equation}
[a_+, a_{+'}^{\dag}] = 2\pi \delta(\Omega-\Omega')\;,\quad
[a_-, a_{-'}^{\dag}] = 2\pi \delta(\Omega-\Omega')\;.
\end{equation}
This leads to the commutation relations for the field amplitudes of quadrature modes 
\begin{eqnarray}
[a_1, a_{2'}^{\dag}] &=& - [a_2, a_{1'}^{\dag}]\; 
= i2\pi\delta(\Omega-\Omega') \nonumber \\
\;[a_1,a_{1'}] &=& [a_1,a_{1'}^{\dag}] = [a_1^{\dag}, a_{1'}^{\dag}] 
= [a_1,a_{2'}] = [a_1^{\dag},a_{2'}^{\dag}] = 0 \;. \nonumber \\
&& 
\end{eqnarray}
\hspace{3mm}As well as (\ref{eqn:26}), the output field can be written as
\begin{eqnarray}
E_{\rm out} 
&=& \sqrt{4\pi\hbar\omega_o\over{\cal A}c} \left[
\cos(\omega_o t)
\int_0^\infty \left(b_1 e^{-\i\Omega t} + b_1^{\dag} e^{+i\Omega t}
\right) {d\Omega\over2\pi}\right. \nonumber\\
&&\quad\quad + \left. \sin(\omega_o t)
\int_0^\infty \left(b_2 e^{-\i\Omega t} + b_2^{\dag} e^{+i\Omega t}
\right) {d\Omega\over2\pi} \right]\;. \nonumber \\
&&
\end{eqnarray}
\hspace{3mm}Then, the relation between the input and output fields with no losses in any optics is \cite{bib:1}, 
\begin{eqnarray}
b_1 &=& a_1 e^{2i\beta} \\
b_2 &=&  a_2 e^{2i\beta} -K a_1 e^{2i\beta} + \sqrt{2K} \left( \frac{h}{h_{SQL}} \right)  e^{i\beta}\;.
\label{eqn:7}
\end{eqnarray}
where various quantities are defined as below.
\begin{eqnarray}
\gamma &= & \frac{T^2c}{4L}\;, \label{eqn:34}\\
&&  \nonumber \\
\beta &=&  \arctan(\Omega / \gamma )\;, \label{eqn:35} \\
&& \nonumber \\
I_{\rm SQL} &=& {mL^2 \gamma^4 \over 4 \omega_o},\;\;\;\;\;\;\;\;\;\;\;\;\;\;\;\;\;\;\;\;\; \\
\nonumber\\
K &=&   \frac{2( I_0/I_{\rm SQL} )}{(\Omega /\gamma )^2 [1+(\Omega /\gamma )^2 ]}  , \quad  \label{eqn:201} \\
\nonumber\\
h_{\rm SQL} &=& \sqrt{8\hbar\over m\Omega^2 L^2} \label{eqn:36}\;.
\end{eqnarray}
\begin{table}
\caption{\label{tab:1}List of several parameters (typical values) of differential-type interferometers.}
\begin{ruledtabular}
\begin{tabular}{lc}
parameter & value \\
\hline
Laser power at a beam splitter & $I_0$ \\
Laser frequency & $\omega_0 \approx 1.8 \times 10^{15}\;\mathrm{s}^{-1}$ \\
Sideband frequency & $\Omega$ \\
Mirror mass & $m=30 \;\mathrm{kg}$ \\
FP cavity's arm length & $L=3 \;\mathrm{km}$ \\
Transmissivity of FP cavity's front mirror  & $T=0.14$ \\
Cavities' halfband width & $\gamma \approx 490 \;\mathrm{s}^{-1}$ \\
Laser power to reach SQL at $\Omega =\gamma$ & $I_{SQL}=2162\;\mathrm{W}$ \\
Amplitude transmissivity of SR mirror  & $\rho=0.98$ \\
detuned phase in the SR cavity & $\phi$ \\
detuned phase in the dark port cavity & $\theta$ \\
\end{tabular}
\end{ruledtabular}
\end{table}
Various parameters we use in this paper are listed in Table \ref{tab:1}. ; $T$ is the amplitude transmissivity of FP cavity's front mirror \cite{bib:14}. $\gamma$ is FP-cavities' half bandwidth, which determines the characteristic frequency of the FP cavity, and $\beta$ is the effective phase shift of a sideband field in the FP cavity. $K$ is a coupling constant between a carrier field and a sideband field, which determines the intensity of radiation-pressure. $h_{SQL}$ is the square root of the SQL spectral density and $I_{SQL}$ is the laser power required to reach SQL at $\Omega=\gamma$. In equation (\ref{eqn:7}), the first term is shot noise, the second term is radiation-pressure noise, and the third term is the GW signal.\\
\hspace{3mm}Converting the noise signal to GW amplitude, $h_n$ is defined as
\begin{equation}
h_n(\Omega ) = \frac{h_{SQL}}{\sqrt{2K}}(a_2-K a_1)e^{i\beta}\;.
\end{equation}
Then, spectral density is defined as the variance of the reduced noise amplitude by   
\begin{equation}
{1\over2}2\pi\delta(\Omega-\Omega')S_h(\Omega) \equiv \langle {\rm in}| h_n(\Omega)
h_n^{\dag}(\Omega')|{\rm in}\rangle_{\rm sym}\;,
\label{eqn:1}
\end{equation}
where subscript "sym" means calculating by replacing $h_n(\Omega) h_n^{\dag}(\Omega')$ with ${1\over2}\left(
h_n(\Omega) h_n^{\dag}(\Omega') + h_n^{\dag}(\Omega') h_n(\Omega)\right)$. $|{\rm in}\rangle $ is an input state. In our configuration, input state at the dark port is in its vacuum state, defined using annihilation operators for each sideband by 
\begin{equation}
a_+|0_a\rangle =a_-|0_a\rangle =0\;.
\end{equation}
Using the relation
\begin{equation}
\langle 0_a|a_j {a}_{k'}^{\dag}|0_a\rangle_{\rm sym} = {1\over 2}
2\pi\delta(\Omega-\Omega')\delta_{jk}\;,
\label{eqn:2}
\end{equation}
we can obtain the spectral density of a conventional recombined-type interferometer
\begin{equation}
S_h = {h^2_{\rm SQL}\over 2}\left( {1\over K} + {K} \right)\;.
\end{equation}
This spectral density reaches SQL at $\Omega = \gamma$ when the laser power $I_0$ is $I_{SQL}$, but does not overcome SQL. The reason is that shot noise and radiation pressure noise have no dynamical correlation. They are proportional and inversely proportional to the laser power, respectively and this gives the achievable minimum noise level.\\

\subsection{Signal-Recycling}
\begin{figure}[t]
\begin{center}
\includegraphics[width=8cm]{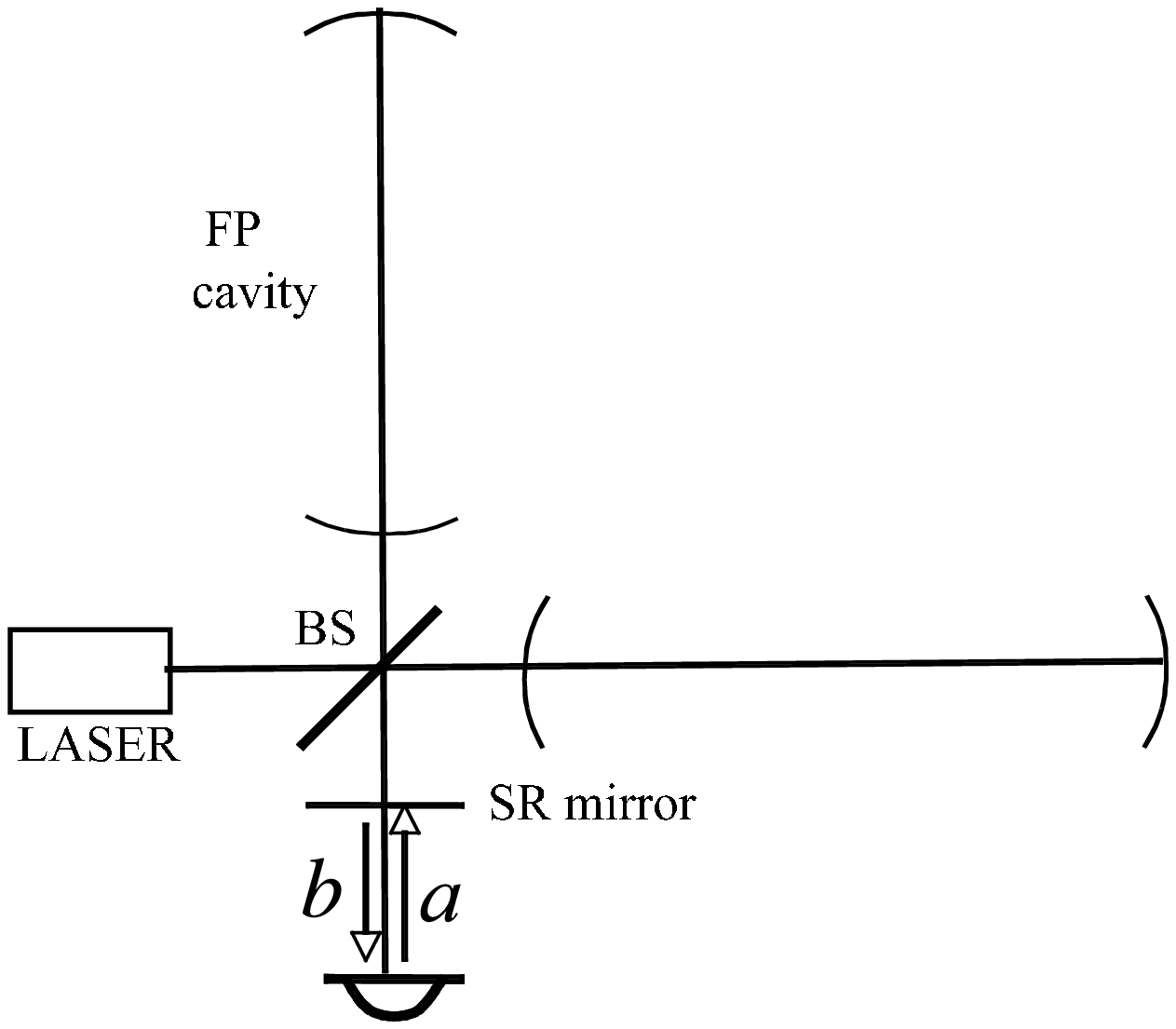}
\caption{SR configuration of recombined-type interferometer.}
\label{fig:3}
\end{center}
\end{figure}
Signal recycling in a recombined-type GW interferometer has been investigated by Buonanno and Chen \cite{bib:2,bib:3}. The SR mirror is located at the dark port (Fig.\ref{fig:3}). Then, the outgoing signal from the beam splitter is reflected by the SR mirror and reenters the interferometer with some phase shift in the SR cavity. The signal circulates in the interferometer many times and creates the resonances at certain frequencies. As a result, the sensitivity has dips at the resonant frequencies.\\
\hspace{3mm}The input-output relation in this configuration is given by \cite{bib:2}
\begin{eqnarray}
\left (
\begin{array}{c}
\displaystyle
b_1 \\
\displaystyle
b_2
\end{array}
\right) &=&
\frac{1}{M}\left[e^{2i(\beta+\Phi)}\left(
\begin{array}{cc}
\displaystyle
C_{11} & C_{12} \\
\displaystyle
C_{21} & C_{22}
\end{array}
\right)
\left(
\begin{array}{c}
\displaystyle
a_1 \\
\displaystyle
a_2
\end{array}
\right) \right. \nonumber \\
&& \left. \;\;\;\;\;\;+
\sqrt{2K}\tau
e^{i(\beta+\Phi)}
\left(
\begin{array}{c}
\displaystyle
D_1 \\
\displaystyle
D_2
\end{array}
 \right)\frac{h}{h_{\rm SQL}}\right], \label{eqn:33} \\
 && \nonumber \\
M & \equiv & 1 + \rho^2\, e^{4i(\beta+\Phi)} \nonumber \\
&&-2\rho\,e^{2i(\beta+\Phi)}\left (\cos{2\phi}+\frac{K}{2}\,
\sin{2\phi} \right )\,,\label{eqn:42} \nonumber  
\end{eqnarray}
\begin{eqnarray}
C_{11} &=& C_{22} \nonumber \\
&\equiv &(1+\rho^2)\,
\left (\cos{2\phi}+\frac{K}{2}\,\sin{2\phi} \right ) -2\rho\,\cos[2(\beta+\Phi)]\,,\nonumber \\
&& \\
&& \nonumber \\
 C_{12} &\equiv & -\tau^2\,(\sin{2\phi}+K\,\sin^2{\phi})\,,  \\
&& \nonumber \\
C_{21} &\equiv& \tau^2\,(\sin{2\phi}-K\,\cos^2{\phi})\,,\\
&& \nonumber \\
D_1 &\equiv & - (1+\rho\, e^{2 i(\beta+\Phi)})\,\sin{\phi}\,, \\
&& \nonumber \\ 
D_2&\equiv& - (-1+\rho\, e^{2 i(\beta+\Phi)})\,\cos{\phi} \;,
\end{eqnarray}
where $\phi \equiv [\omega _0 \ell /c]_{\mod 2\pi}$ is the phase shift in the SR cavity for the carrier field and $\Phi \equiv [\Omega _0 \ell /c]_{\mod 2\pi}$ is the phase shift in the SR cavity for the sideband field. $\ell$ is the length between SR mirror and the beam splitter. $\Phi$ can be ignored because $\ell$ is typically of the order of several meters. As we can see from (\ref{eqn:33}), GW signals appear in both quadrature modes. Thus, homodyne detection angle $\zeta$ is also an important parameter and in general, the output signal is written as
\begin{equation}
b_\zeta =b_1\,\sin\zeta+b_2\,\cos \zeta \;.
\end{equation}
From the above input-output relation, the spectral density is derived as
\begin{eqnarray}
S_h&=& \frac{h_{\rm SQL}^2}{2K\tau^2} \nonumber \\
&\times& \frac{\left(C_{11}\,\sin\zeta+C_{21}\,\cos\zeta\right)^2+
\left(C_{12}\,\sin\zeta+C_{22}\,\cos\zeta\right)^2}
{\left|D_1\,\sin\zeta+D_2\,\cos\zeta\right|^2}\,.\nonumber \\
&&
\end{eqnarray}
\hspace{3mm}One can verify that this spectral density has at most two dips on the sensitivity curve (see \cite{bib:2}). These two dips have different origins. One corresponds to an optical resonance (sideband resonance) and the other corresponds to a mechanical resonance (optical rigidity). The optical resonance is just the resonance in the cavities due to the sideband fields. On the other hand, the mechanical resonance has the origin in the suspension system. When the laser power is high, the resonant frequency is shifted upward into the detection band by the optical rigidity,
which is caused by the nontrivial coupling between radiation pressure and mirror motion. In the case with detuned SR phase $\phi$, the mirror no longer behaves like a free mass, but like a mass attached to a mechanical spring due to optical fields. These resonances create two dips in the sensitivity curve and one can reach the sensitivity beyond the SQL.\\
 
\section{Quantum noise in a Differential-type interferometer}
\begin{figure}[t]
\begin{center}
\includegraphics[width=8cm]{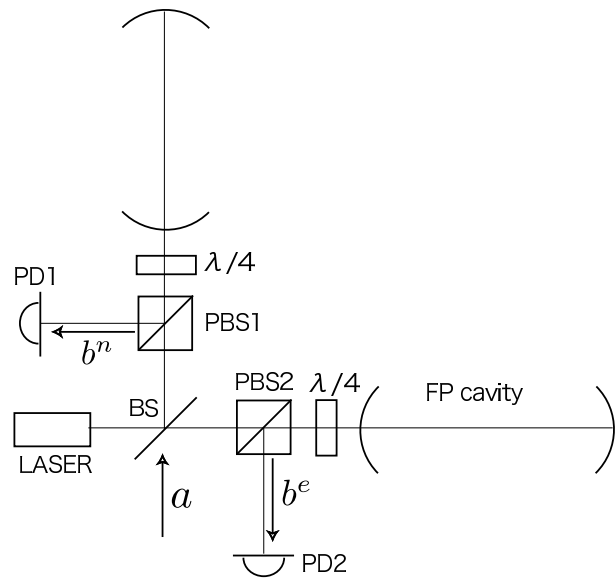}
\caption{Conventional configuration of differential-type interferometer, and input and output fields.}
\label{fig:1}
\end{center}
\end{figure}
\subsection{conventional interferometer}
The differential-type interferometer we consider is shown in Fig.\ref{fig:1}. Laser light entering the beam splitter (BS) is split into two directions and enters a pair of a polarization beam splitter (PBS) and a quarter wavelength ($\lambda/4$) plate
at each arm \cite{bib:17}. Only the light with horizontal polarization is transmitted through PBS and $\lambda/4$ plate and goes into the FP cavity. After being reflected by the FP cavity, the light is transmitted through the $\lambda/4$ plate and is reflected by PBS since it has vertical polarization. Then the beams are detected at the photo detectors independently in each arm.\\
\hspace{3mm}For simplicity, we assume that all optics (beam splitter, PBS, $\lambda/4$ plate and the mirrors of FP cavity) are lossless. The beam splitter has a reflectivity of 0.5. The end mirrors of the FP cavity are completely reflective and its front mirrors have amplitude transmissivity and reflectivity $\{+T, -R\}$ for ingoing light and $\{+T, +R\}$ for outgoing light. $T^2+R^2=1$ is satisfied since we assume the mirrors are lossless. The zeroth order length of the FP cavity satisfies resonant condition $L=n\lambda/2$ ; $n$ is the integer and $\lambda$ is the wavelength of carrier light. Other lengths of the light path in an interferometer (between beam splitter and PBS, between photo detector and PBS, and between PBS and a front mirror of FP cavity) are set to make no phase shift for carrier light and are small enough to make only negligible phase shift for a sideband field. \\
\hspace{3mm}"$a$" is a vacuum field (input field in this configuration) which enters the beam splitter from the dark port. "$b^n\;$" and "$b^e\;$" are output fields. Subscripts "n" and "e" denote "north" and "east" respectively. It should be noticed that the vacuum fluctuation we should consider, coming into this interferometer, is only "$a$" because other vacuum fields have polarizations different from that of the field we are interested in. The details of the reason why the other vacuum fields do not contribute to noise are described in Appendix. Treating vacuum fields quantum-mechanically, we can derive the following relation between input and output fields. The detailed calculation is described in Appendix and the results are
\begin{eqnarray}
\Delta b_1 &\equiv & b_1^n -b_1^e \nonumber \\
&=& \sqrt{2} e^{2 i \beta} a_1 \\
\Delta b_2 &\equiv & b^n_2-b^e_2\nonumber \\
&=& \sqrt{2} e^{2i\beta} ( a_2 -K a_1 )+2 \sqrt{K} \left( \frac{h}{h_{SQL}} \right) e^{i\beta}
\end{eqnarray}
where $\Delta b_i,\;i=1,2$ is a differential signal between the signal for each arm in the Fourier domain and $\beta, K, h_{SQL}$ are defined by (\ref{eqn:35}), (\ref{eqn:201}), and (\ref{eqn:36}) in the previous section. This is the same formula as that for the recombined-type except for an overall factor $\sqrt{2}$, which appears because the signals are detected in each arm before being split at the beam splitter. The overall factor does not affect the spectral density. Thus, we obtain the same spectral density as that in a recombined-type. It does not overcome SQL, but can reach SQL at $\Omega =\gamma$ with $I_0=I_{SQL}$. It should be noticed again that this spectral density has no correlation between shot noise and radiation pressure noise. However, the problem is that $I_{SQL}$ is a little large to reach SQL ; $I_{SQL}\approx 2160\;\mathrm{W}$ for the parameters listed in Table \ref{tab:1}. This is a fatal defect for a conventional differential-type interferometer because it is impossible to implement power recycling. Therefore, there seems to be no advantage in using a conventional differential-type interferometer instead of a conventional recombined-type one, from the point of view of the sensitivity.\\

\subsection{Signal-Recycling interferometer}
\begin{figure}[h]
\begin{center}
\includegraphics[width=8cm]{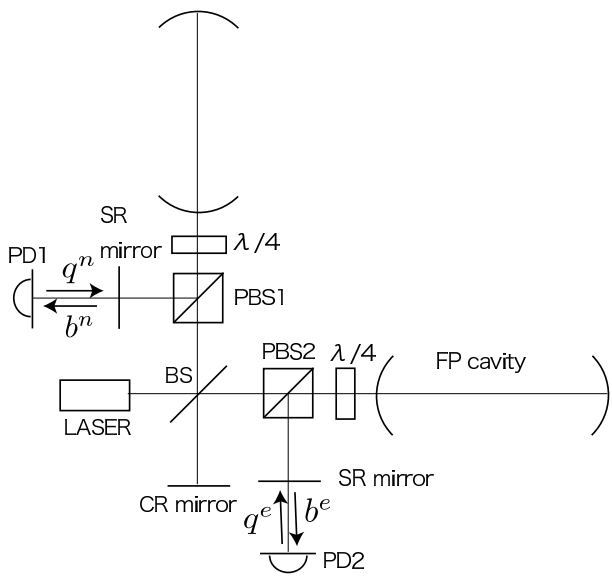}
\caption{SR configuration of a differential-type interferometer and input and output fields.}
\label{fig:2}
\end{center}
\end{figure}
\begin{figure}[t]
\begin{center}
\includegraphics[width=8cm]{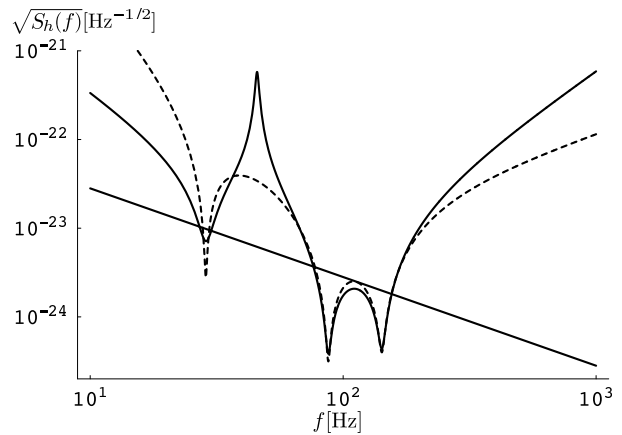}
\caption{Characteristic sensitivity curve of differential-type SR interferometer. Parameters selected are $T=0.14$, $\rho=0.98$, $I_0=I_{SQL}$, $\phi=1.4$, $\theta=0.86$. Solid curve is the sensitivity of quadrature mode 1 ($\zeta=\pi/2$) and dashed curve is that of quadrature mode 2 ($\zeta=0$). Diagonal black line is $h_{SQL}$.}
\label{fig:4}
\end{center}
\end{figure}

Next, we will consider the signal-recycling configuration of a differential-type interferometer. To implement this, SR mirrors should be put just in front of the photo detectors.  We also need to put completely reflecting mirror (CR mirror) at the dark port in order to close the entire system of the interferometer and to recycle outgoing signals. Then, no vacuum field comes into the interferometer from outside the dark port, and only $\mathbf{q}$ fields, which are the vacuum fields coming from the photo detectors in this SR configuration, contribute to noise. These fields do not affect noise in the case without SR mirrors. However, in this SR configuration, $\mathbf{q}$ field couples with the carrier light reflected by the SR mirror in the FP cavity and creates radiation pressure fluctuation. In addition, $\mathbf{q}$ field reflected at CR mirror, reenters the FP cavity, then couples with the carrier light and also creates radiation-pressure fluctuation. Therefore, this interferometer configuration makes the behavior of the sideband fields very complicated. \\
\hspace{3mm}This interferometer configuration has two important parameters $\phi$ and $\theta$,
\begin{equation}
\phi \equiv  \left[ \frac{\omega_0 \ell_s}{c}\right] _{\mod\;2\pi},\;\;\;\;\;\;\theta \equiv \left[ \frac{\omega_0 \ell_d}{c} \right] _{\mod\;2\pi}\;,  
\end{equation}
where $\ell_s$ and $\ell_d$ are the distances of the SR cavity and the dark port cavity. More strictly, these are the distances between the PBS and the SR mirror and between the beam splitter and the CR mirror, respectively. We assume these lengths are small compared with the FP cavity's arm length $L$ ($\ell_s,\; \ell_d\;\sim$ several meters ). Thus, phase shifts for sideband in these cavities are negligible and we will ignore them hereafter. Adjusting these parameters, the shape of the interferometer noise curve changes significantly. $\phi$ has to be set to the same value for both arms, otherwise the common mode of noise signal contributes to the final differential signal and worsens the sensitivity. 
Compared with the recombined-type one, one parameter is added (recombined-type has only "$\phi$" as a parameter). This produces the more variations of noise curve than that of a recombined-type, though it also makes the behavior of the system more complicated.\\
\hspace{3mm}Amplitude reflectivity and transmissivity of the SR mirror are defined as ${-\rho,+\tau}$ for the light entering SR mirror from outside the interferometer and as ${+\rho,+\tau}$ for the light entering SR mirror from inside. These satisfy $\rho^2+\tau^2=1$. \\
\hspace{3mm}The input-output relation of this configuration can be derived after lengthy but straightforward calculations described in the Appendix, becoming 
\begin{widetext}
\begin{eqnarray}
\Delta \mathbf{b}&=&\frac{1}{M}\left[ e^{4i\beta} 
\left(
\begin{array}{cc} \displaystyle
C_{11} & C_{12} \\
\displaystyle
C_{21} & C_{22}
\end{array}\right)
\Delta \mathbf{q}
+2 \tau \sqrt{K}e^{i\beta}
\left(
\begin{array}{c} \displaystyle
D_1  \\
\displaystyle
D_2 
\end{array}\right)
\left( \frac{h}{h_{SQL}} \right)
\right]
\label{eqn:3}\\
&& \nonumber \\
M&=&1+\rho^2 e^{8i\beta} \nonumber \\
&&-2\rho \;e^{4i \beta} \left[ \cos2(\theta +\phi )+\frac{K}{2} \left\{ (1+\rho ^2)\;\sin2(\theta +\phi )+(e^{-2i\beta}+\rho ^2 e^{2i\beta})\;\sin2\theta +2\rho \;\cos2\beta\; \sin2\phi \right\} \right]  \\
&& \nonumber 
\end{eqnarray}
\\
\begin{eqnarray}
C_{11}&=& (1+\rho^2 )\;\cos2(\theta +\phi )-2\rho \;\cos4\beta \nonumber \\
&&+ \frac{K}{2}\left[ (1+\rho ^2)^2\;\sin2(\theta +\phi ) -\tau ^4 \;\sin2\theta + 2\rho \; \cos2\beta \{ (1+\rho ^2)\; \sin2\phi +2\rho \; \sin2\theta \} \right] \label{eqn:37} \\
&& \nonumber \\
C_{22}&=& (1+\rho^2 )\;\cos2(\theta +\phi )-2\rho \;\cos4\beta \nonumber \\
&&+ \frac{K}{2}\left[ (1+\rho ^2)^2\;\sin2(\theta +\phi ) +\tau ^4 \;\sin2\theta + 2\rho \; \cos2\beta \{ (1+\rho ^2)\; \sin2\phi +2\rho \; \sin2\theta \} \right] \\ 
&& \nonumber \\
C_{12}&=&-\tau ^2 \left[ \sin2(\theta +\phi )+K \; \sin\phi \; \{ (1+\rho ^2)\;\sin(2\theta +\phi )+ 2\rho \; \cos2\beta \; \sin\phi \} \right] \\
&& \nonumber \\
C_{21}&=&\tau ^2 \left[ \sin2(\theta +\phi )-K \; \cos\phi \; \{ (1+\rho ^2)\;\cos(2\theta +\phi )+ 2\rho \; \cos2\beta \; \cos\phi \} \right] \\
&& \nonumber \\
D_1 &=& -\left[ (1+\rho^2 e^{6i\beta})\;\sin\phi +2\rho \;e^{3i\beta}\;\cos\beta \;\sin(2\theta +\phi ) \right] 
\label{eqn:202} \\
&& \nonumber \\
D_2 &=& -\left[ (-1+\rho^2 e^{6i\beta})\;\cos\phi +2i\rho \;e^{3i\beta}\;\sin\beta \;\cos(2\theta +\phi ) \right] \;.
\label{eqn:203} 
\end{eqnarray}
\end{widetext}
$C_{ij}$ are real coefficients and involve the contribution of quantum noise and $D_{i}$ are complex coefficients and involve GW signal amplification or suppression. In the above equations, an interesting feature appears. The effective phase shift in the FP cavity doubles compared with that of the recombined-type because the differential-type SR interferometer effectively has two FP cavities when light goes around the interferometer and this extends the length of the light path in the interferometer.\\
\hspace{3mm}It is straightforward to calculate spectral density from input-output relation as well as the previous section. In general, we assume homodyne detection with angle $\zeta$. Thus, the differential signal becomes
\begin{eqnarray}
\Delta b_{\zeta} &\equiv & \Delta b_1 \;\sin\zeta +\Delta b_2 \;\cos\zeta \nonumber \\
&=&(b_1^n \;\sin\zeta +b_2^n \;\cos\zeta )-(b_1^e \;\sin\zeta +b_2^e \;\cos\zeta )\;. \nonumber \\
\end{eqnarray}
Now, the input state is in a vacuum state and is defined as
\begin{equation}
q^n_+|0_q\rangle =q^n_-|0_q\rangle =q^e_+|0_q\rangle =q^e_-|0_q\rangle =0 \;.
\label{eqn:4}
\end{equation}
Then, using (\ref{eqn:1}) and (\ref{eqn:2}), spectral density is given by the following equation.
\begin{eqnarray}
S_h &=& {h^2_{\rm SQL}\over 2 K \tau^2} \nonumber \\
&\times & \frac{(C_{11}\sin\zeta +C_{21}\cos\zeta )^2+(C_{12}\sin\zeta +C_{22}\cos\zeta )^2}{|D_1\sin\zeta +D_2\cos\zeta |^2} \nonumber \\
&&
\label{eqn:16}
\end{eqnarray}
This is the same formula as that of a recombined-type signal recycling, but the coefficients are different.\\
\hspace{3mm}A characteristic sensitivity curve of a differential-type SR interferometer for certain parameters is drawn in Fig.\ref{fig:4}. There appears three dips and one GW signal suppression peak, which we will discuss in detail in the next section. In comparison with a recombined-type, there appears one additional dip. The important thing here is that these three dips overcome the SQL.\\ 

\section{decomposition of spectral density and physical interpretaion}
The above expression of spectral density is rather complicated to analyze and to interpret its physical meaning. So, in this section, we will use the linear quantum
measurement theory and decompose spectral density into three parts: shot noise, radiation-pressure noise, and correlation noise \cite{bib:9,bib:3}. \\
\subsection{the decomposition of spectral density}
First, we have to decompose the input-output relation into shot noise and radiation-pressure parts. Devoting our attention to the dependence on mirror mass $m$, normalized input-output relation is given by 
\begin{equation}
{\cal{O}}_i(\Omega )={\cal{Z}}_i(\Omega )+R_{xx}(\Omega ){\cal{F}}_i(\Omega )+Lh(\Omega )
\end{equation}
This is the equation divided (\ref{eqn:3}) by certain coefficients to give the displacement of a mirror. ${\cal{Z}}_i$ is shot noise, which is independent of $m$. ${\cal{F}}_i$is radiation-pressure noise, which scales proportional to $m^{-1}$. $R_{xx}$ is the susceptibility of the asymmetric mode of motion of FP cavity's mirrors and is defined as $R_{xx}(\Omega ) \equiv -4/(m\Omega ^2)$. Specific forms of these for each quadrature mode are,
\begin{widetext} 
\begin{eqnarray}
{\cal{Z}}_1(\Omega ) &=&\frac{Lh_{SQL}}{2\tau \sqrt{K} D_1}\;e^{3i\beta} \left[ \{ (1+\rho ^2)\;\cos2(\theta +\phi )-2\rho \; \cos4\beta \} \Delta q_1 -\tau ^2\;\sin2(\theta + \phi ) \Delta q_2 \right] \\
\nonumber \\ 
{\cal{Z}}_2(\Omega ) &=&\frac{Lh_{SQL}e^{3i\beta}}{2\tau \sqrt{K} D_2}\;e^{3i\beta} \left[ \{ (1+\rho ^2)\;\cos2(\theta +\phi )-2\rho \; \cos4\beta \} \Delta q_2 +\tau ^2\;\sin2(\theta + \phi ) \Delta q_1 \right]\\
\nonumber \\
{\cal{F}}_1(\Omega ) &=&\frac{Lh_{SQL}\sqrt{K}}{4\tau R_{xx} D_1} \;e^{3i\beta} \left[ \{ (1+\rho ^2)^2\;\sin2(\theta +\phi )+2\rho \;\cos2\beta ((1+\rho ^2)\;\sin2\phi +2\rho \;\sin2\theta )-\tau ^4 \;\sin2\theta  \} \Delta q_1 \right. \nonumber \\
&& \left. -2\tau ^2 \sin\phi \{ (1+\rho ^2)\;\sin(2\theta +\phi )+2\rho \; \cos2\beta \;\sin\phi  \} \Delta q_2 \right]\\
\nonumber \\ 
{\cal{F}}_2(\Omega )  &=&\frac{Lh_{SQL}\sqrt{K}}{4\tau R_{xx} D_2} \;e^{3i\beta} \left[ \{ (1+\rho ^2)^2\;\sin2(\theta +\phi )+2\rho \;\cos2\beta ((1+\rho ^2)\;\sin2\phi +2\rho \;\sin2\theta )+\tau ^4 \;\sin2\theta  \} \Delta q_2 \right. \nonumber \\
&& \left. -2\tau ^2 \cos\phi \{ (1+\rho ^2)\;\cos(2\theta +\phi )+2\rho \; \cos2\beta \;\cos\phi  \} \Delta q_1 \right]
\end{eqnarray}
\end{widetext}
where $D_1$ and $D_2$ are given by (\ref{eqn:202}) and (\ref{eqn:203}). Using (\ref{eqn:2}) and the definition of spectral density
\begin{eqnarray}
\frac{1}{2}S_{AB}(\Omega )2\pi \delta (\Omega -\Omega ^{\prime})&& \nonumber \\
=\frac{1}{2} \langle A(\Omega )B^{\dag}(\Omega ^{\prime})&+&B^{\dag}(\Omega ^{\prime})A(\Omega ) \rangle,
\end{eqnarray}
and calculating spectral density of each noise part gives,
\begin{eqnarray}
\bar{S}_{{\cal{Z}}_1{\cal{Z}}_1} &=& \frac{2L^2h_{SQL}^2}{\tau^2 K |D_1|^2} [ \cos2(\theta +\phi )-\cos4\beta ]^2 \label{eqn:27}\\
&& \nonumber \\
\bar{S}_{{\cal{Z}}_2{\cal{Z}}_2} &=& \frac{2L^2h_{SQL}^2}{\tau^2 K |D_2|^2} [ \cos2(\theta +\phi )-\cos4\beta ]^2  \label{eqn:28} \\
&& \nonumber \\
\bar{S}_{{\cal{F}}_1{\cal{F}}_1} &=& \frac{2L^2h_{SQL}^2 K}{\tau^2 R_{xx}^2 |D_1|^2} \nonumber \\
&&\times \left[ \sin2(\theta +\phi )+\cos2\beta \{ \sin2\theta +\sin2\phi \} \right]^2 \nonumber \\
\label{eqn:29} 
\end{eqnarray}
\begin{eqnarray}
\bar{S}_{{\cal{F}}_2{\cal{F}}_2} &=& \frac{2L^2h_{SQL}^2 K}{\tau^2 R_{xx}^2 |D_2|^2} \nonumber \\
&&\times \left[ \sin2(\theta +\phi )+\cos2\beta \{ \sin2\theta +\sin2\phi \} \right]^2 \nonumber \\
\label{eqn:30} 
\end{eqnarray}
\begin{eqnarray}
\bar{S}_{{\cal{Z}}_1{\cal{F}}_1} &=& S_{{\cal{F}}_1{\cal{Z}}_1} \nonumber \\
&=& -\frac{2L^2h_{SQL}^2}{\tau^2 R_{xx} |D_1|^2}  [\cos2(\theta +\phi )-\cos4\beta ] \nonumber \\
&&\times [\sin2(\theta +\phi )+\cos2\beta \{ \sin2\theta +\sin2\phi \}] \nonumber \\
\label{eqn:31}  
\end{eqnarray}
\begin{eqnarray}
\bar{S}_{{\cal{Z}}_2{\cal{F}}_2} &=& S_{{\cal{F}}_2{\cal{Z}}_2} \nonumber \\
&=& -\frac{2L^2h_{SQL}^2}{\tau^2 R_{xx} |D_2|^2}  [\cos2(\theta +\phi )-\cos4\beta ] \nonumber \\
&&\times [\sin2(\theta +\phi )+\cos2\beta \{ \sin2\theta +\sin2\phi \}] \nonumber \\
\label{eqn:32} 
\end{eqnarray}
Here we assumed that the SR mirror is highly reflective and used the approximation taking the leading terms about $\tau$. This is not a strong constraint for practical purposes because we want to use highly reflective SR mirrors to implement good sensitivity. Moreover, this approximation makes it easy to interpret the physical behavior of the system.\\
\hspace{3mm}Total spectral density is given simply by adding three noise parts.
\begin{equation}
\bar{S}_h \approx \frac{1}{L^2}[\bar{S}_{{\cal{Z}}{\cal{Z}}}+R_{xx}^2\bar{S}_{{\cal{F}}{\cal{F}}}+2R_{xx}\bar{S}_{{\cal{Z}}{\cal{F}}}]
\end{equation}

\subsection{Number of dips and their resonant frequency}
We will evaluate the number of dips and their positions in a sensitivity curve in this subsection. There are two kinds of resonances in a SR interferometer; optical resonance (sideband resonance) and mechanical resonance (optical rigidity) as we explained in Chapter II B. We will describe details below.\\
\hspace{3mm}Optical dips correspond to the resonances that certain sideband fields resonates in the interferometer. It appears even when the terms of the spectral density concerning radiation-pressure is negligible, $\bar{S}_{{\cal{F}}_i{\cal{F}}_i} \rightarrow 0$ and $\bar{S}_{{\cal{Z}}_i{\cal{F}}_i} \rightarrow 0$ for $i=1,2$. Thus, the frequency of optical dips made by shot noise can be calculated as the solutions of the equation $\bar{S}_{{\cal{Z}}_i{\cal{Z}}_i}=0,\;i=1,2$. Note that, as long as the leading term of the spectral density about $\tau$ is considered, both quadrature modes give same equations. Rewriting the equation with the following relations,
\begin{eqnarray}
y &\equiv & \left( \frac{\Omega_{res}}{\gamma } \right) ^2, \nonumber \\
\cos4\beta &=&\frac{1-6y+y^2}{(1+y)^2}, \nonumber
\end{eqnarray}
gives
\begin{equation}
1-6y+y^2=(1+y)^2 \;\cos2(\theta +\phi ),
\label{eqn:14}
\end{equation}
where $\Omega_{res}$ is the angular frequencies of the resonances. The solutions are
\begin{equation}
y_s=\frac{3+\cos2(\theta +\phi ) \pm 2\sqrt{2\{ 1+\cos2(\theta +\phi ) \}}}{1-\cos2(\theta +\phi )}\;.
\label{eqn:20}
\end{equation}
As one can see from the solutions (\ref{eqn:20}), $y_s$ are real solutions and have two resonant frequencies for all $\phi$ and $\theta$ parameters except for the case $\phi +\theta = (2n+1)\pi/2\;,\;n=integer$.\\
\hspace{3mm}When the laser power is high, above resonant conditions $\bar{S}_{{\cal{Z}}_i{\cal{Z}}_i}=0,\;i=1,2$ are no longer valid and the optical resonant frequencies are shifted due to the effect of radiation pressure. Moreover, correlation part of spectral density makes one more dip. The position of a mechanical dip is determined by the equation, $\bar{S}_{{\cal{Z}}_i{\cal{Z}}_i}+R_{xx}^2\bar{S}_{{\cal{F}}_i{\cal{F}}_i}+2R_{xx}\bar{S}_{{\cal{Z}}_i{\cal{F}}_i}=0,\;i=1,2$. Using the relations,
\begin{eqnarray}
\cos2\beta &=&\frac{1-y}{1+y} \nonumber \\
K&=&2n/y/(1+y) \nonumber \\
n&\equiv & I_0/I_{SQL} \nonumber \;
\end{eqnarray}
and rewriting the resonant equation, we obtain
\begin{eqnarray}
&& y\left[(1+y)^2 \;\cos2(\theta +\phi )-( 1-6y+y^2) \right] \nonumber \\
&=&2n \left[ (1+y)\;\sin2(\theta +\phi )+(1-y)\;(\sin2\theta +\sin2\phi ) \right]. \nonumber \\
\;
\label{eqn:206}
\end{eqnarray}
\begin{figure}[t]
\begin{center}
\includegraphics[width=6cm]{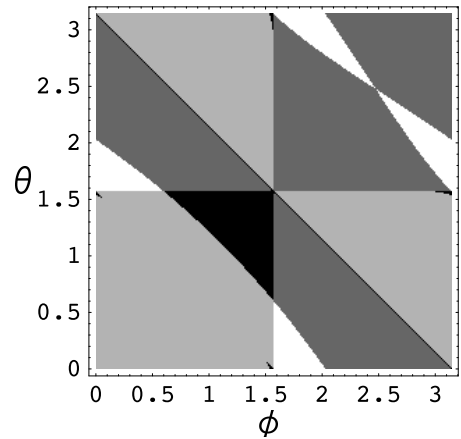}
\caption{Number of dips when $n=1$. We numerically solved (\ref{eqn:206}) and showed the number of dips, which has real frequency, with colors. Black, dark gray, and light gray regions have three, two, and one solution, respectively and white regions have no solution.}
\label{fig:51}
\end{center}
\end{figure}
\begin{figure}[h]
\begin{center}
\includegraphics[width=8cm]{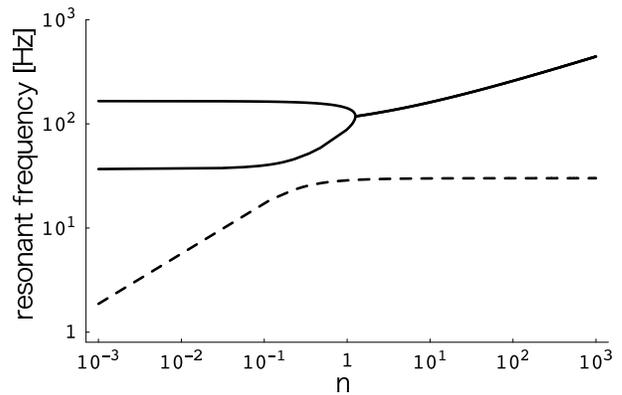}
\caption{Laser power dependence of the positions of resonant frequency for the parameters $\{T=0.14,\; \phi=1.4,\;\theta=0.86 \}$.  We evaluated (\ref{eqn:206}) and showed the resonant frequencies with solid line for optical resonance and dashed line for mechanical resonance.}
\label{fig:52}
\end{center}
\end{figure}

\begin{figure*}[t]
\begin{center}
\includegraphics[width=12cm]{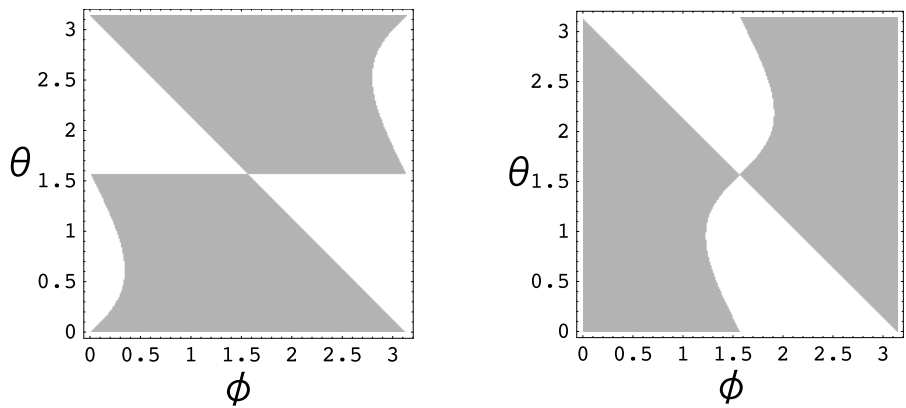}
\caption{Number of solutions of GW suppression. We evaluated (\ref{eqn:22}) and (\ref{eqn:25}), and showed whether the solution exists or not with colors. Shaded regions have one solution and white regions have no solution. The left panel shows that of the suppression frequency for quadrature mode 1 and the right panel is for quadrature mode 2.}
\label{fig:6}
\end{center}
\end{figure*}
\begin{figure*}[t]
\begin{center}
\includegraphics[width=17cm]{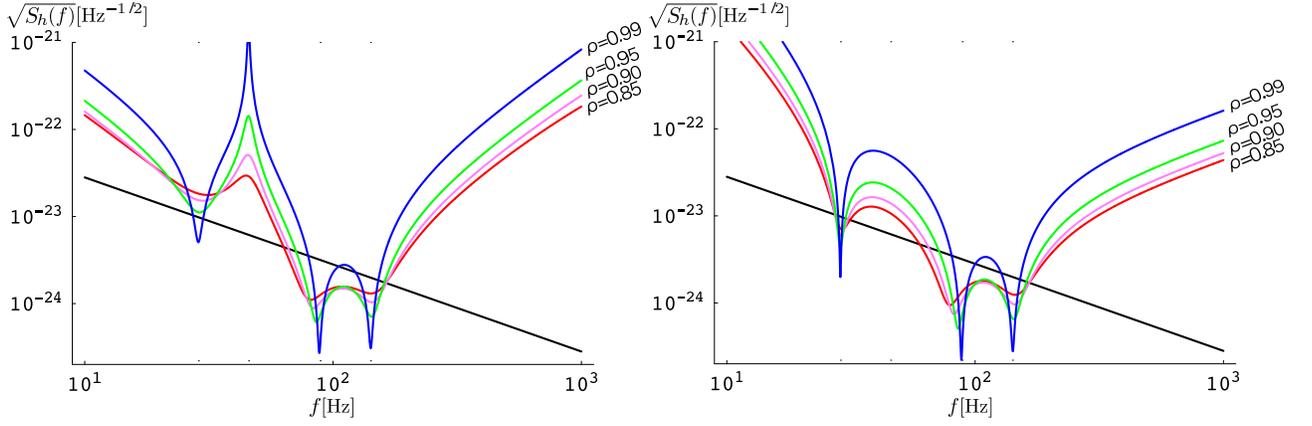}
\caption{(color online). Discrepancy of resonant frequencies with/without the approximation, and the dependence on the reflectivity of the SR mirror $\rho$. Left and right panels show the cases of $\zeta=\pi/2$ (quadrature 1) and $\zeta=0$ (quadrature 2). Other parameters are selected as $T=0.14$, $I_0=I_{SQL}$, $\phi=1.4$, $\theta=0.86$. Vertical lines are the dip frequencies predicted by the approximated solutions. Each curve is drawn without the approximation and with $\rho=0.99$, $\rho=0.95$, $\rho=0.90$, $\rho=0.85$, respectively as indicated in the figure. As $\rho$ is smaller, the dips become shallower and the positions of the dips are slightly shifted. However, the discrepancy is not significant and is within the order of ${\cal{O}}(\tau^2)$.
}
\label{fig:9}
\end{center}
\end{figure*}

In general, this equation has three solutions. Two of three solutions are optical resonances because the left hand side of the resonant equation (\ref{eqn:206}) is the same as the equation (\ref{eqn:14}) for pure optical resonances. On the other hand, one can easily verify that the right hand side of the resonant equation has the same form as $\bar{S}_{{\cal{F}}_i{\cal{F}}_i}=0,\;i=1,2$. The above resonant equation is equivalent to the equation $\sqrt{\bar{S}_{{\cal{Z}}_i{\cal{Z}}_i}}=-R_{xx}\sqrt{\bar{S}_{{\cal{F}}_i{\cal{F}}_i}}$. One can solve this equation analytically, however, the solutions are rather complicated and also depends on the ratio of the laser power $I_0$ and $I_{SQL}$. So, we will evaluate the number of the resonant frequencies and its position on the sensitivity curve numerically for the certain parameters of the interferometer. Selected parameters are listed in Table \ref{tab:1}. Figure \ref{fig:51} shows the number of dips. In general, the solutions of the equation can be complex number. In the case, $(\Omega_{res}/\gamma)$ also has imaginary part. This means $Re[(\Omega_{res}/\gamma)]$ does not satisfy the equation (\ref{eqn:206}) and not make sharp dips. So, we do not count such dips. As shown in Fig.\ref{fig:51}, there are at most three dips in the sensitivity. Figure \ref{fig:52} shows the laser power dependence of the resonant frequencies. In the limit of $n\rightarrow 0$, optical resonance approaches certain frequencies given by (\ref{eqn:20}). On the other hand, one of three solutions approaches $\Omega_{res}=0$. So, we can conclude that the solution is a mechanical dip. \\
\hspace{3mm}Comparison with a recombined-type SR interferometer helps our understanding of the number of dips. The recombined-type has one optical dip and one mechanical dip. However, our differential-type interferometer has one additional dip. The resonant condition of optical dips $\bar{S}_{{\cal{Z}}_i{\cal{Z}}_i}=0,\;i=1,2$ gives $\cos4\beta =\cos2(\phi+\theta)$. This condition is satisfied if $\pm4\beta +2(\phi+\theta) =2 \pi m.\;(m\;:\;integer)$. For simplicity, let $\theta$ be set to zero. The resonant condition has the same form as a recombined-type SR interferometer except for the doubled phase shift $2\beta \rightarrow 4\beta$, c.f. in recombined-type, $\cos2\beta =\cos2\phi$. This is because the light passes through the FP cavity twice when it goes around a differential-type interferometer, for instance, when the light starts at SR mirror and comes back at SR mirror. In other words, two FP cavities are coupled. This allows the sideband field to increase the resonant solution. Strictly speaking, in the case of a recombined-type, two sidebands $\pm \Omega$ satisfy the same condition and have a degenerated resonant frequency, on the other hand, in a differential-type, two sidebands $\pm \Omega$ satisfy asymmetric resonant conditions and have different resonant frequencies. Therefore, a differential-type SR interferometer makes more dips.\\
\hspace{3mm}Such a behavior also has been seen in other detector configurations, for exsample, Sagnac interferometers with two cavities \cite{bib:18,bib:19,bib:20}. In these references, Sagnac interferometers without SR and with tuned-SR have been investigated as a speed meter. There are some common points with our differential-type configuration and, in fact, light is injected into two cavities in a row and obtains the doubled phase shift in the cavity. This is also true in the recombined-type RSE with the long SR cavity though they assume the SR cavity is short and ignore the phase shift of the sideband in it \cite{bib:2}. Therefore, three dips also may appear in these configurations if the detuned SR is done.\\
\hspace{3mm}There also exists uninteresting peak on the noise curve due to GW signal suppression. Mathematically, the denominators of the spectral density approaches zero. The  suppression frequency is determined by the equation $|D_1|=0$ for quadrature mode 1 and $|D_2|=0$ for quadrature mode 2. First, we shall consider quadrature mode 1 ; $|D_1|=0$. From (\ref{eqn:202}), if we take the leading order about $\tau$, it becomes
\begin{equation}
(1+y)[\sin(2\theta +\phi )- \sin\phi ] +2(1-y)\sin\phi =0\;.\nonumber 
\end{equation}
The solution is,
\begin{equation}
y_{GW}^{(1)}=\frac{\sin\phi +\sin(2\theta +\phi )}{3\sin\phi -\sin(2\theta +\phi )}\;.
\label{eqn:22}
\end{equation}
As well, for quadrature mode 2, using (\ref{eqn:203}), $|D_2|=0$ becomes
\begin{equation}
(1+y)[\cos(2\theta +\phi )+ \cos\phi ] +2(1-y)\cos\phi =0\;.\nonumber 
\end{equation}
The solution is,
\begin{equation} 
y_{GW}^{(2)}=\frac{3\cos\phi +\cos(2\theta +\phi )}{\cos\phi -\cos(2\theta +\phi )}\;.
\label{eqn:25}
\end{equation} 
Whether the solutions due to GW signal suppression exist or not is shown in Fig.\ref{fig:6}.\\
\hspace{3mm}We shall summarize the number of dips. The shot noise part has two optical dips, whose resonant frequencies are shifted by the effect of radiation pressure, and the radiation-pressure noise part and the correlation noise part make one mechanical dip in the detection band. Compared with a recombined-type interferometer, a differential-type interferometer has one more additional dip due to the doubled optical path in the interferometer. There also exists uninteresting GW suppression for each quadrature. At the frequency, GW signal is canceled out and the noise curve has a large peak. \\
\hspace{3mm}In the above analysis, resonant frequencies of the system have been obtained under the approximation which takes the leading terms about $\tau$ in each spectral density. We have to make the coverage of the approximation clear. Figure \ref{fig:9} shows the comparison of the resonant frequencies with/without the approximation. When $\rho=0.99$, resonant frequencies show good agreements with the approximated solutions. When the reflectivity of the SR mirror is smaller, the disagreement is larger. The magnitude of the disagreement is ${\cal{O}}(\tau ^2)$. As the figure shows, the resonant frequencies are less sensitive to the reflectivity of the SR mirror, though the depth of dips are very sensitive to it.\\

\section{Application of differential-type SR interferometer to real GW interferometer}
\begin{figure}[t]
\begin{center}
\includegraphics[width=8.5cm]{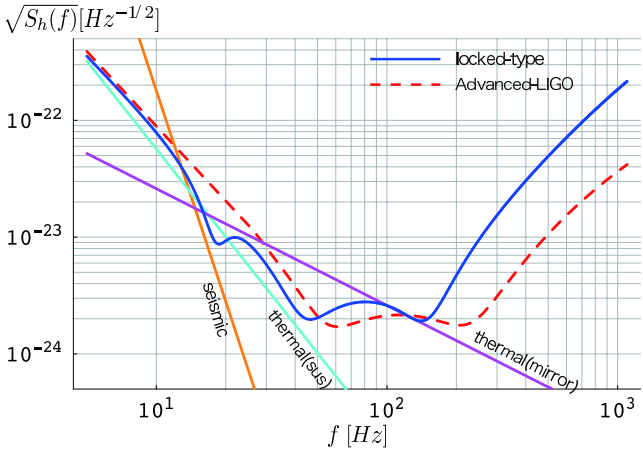}
\caption{(color online). Comparison of the sensitivity curves of a differential-type SR interferometer and Advanced-LIGO. Solid and dashed curve are the sensitivity curve of the differential-type with adjusted parameters listed in Table \ref{tab:2} and Advanced-LIGO. Other straight lines are noises of Advanced-LIGO; thermal noise of a suspension, thermal noise of a mirror and seismic noise as indicated in the figure \cite{bib:10}.}
\label{fig:11}
\end{center}
\end{figure}

\begin{figure}[t]
\begin{center}
\includegraphics[width=8.5cm]{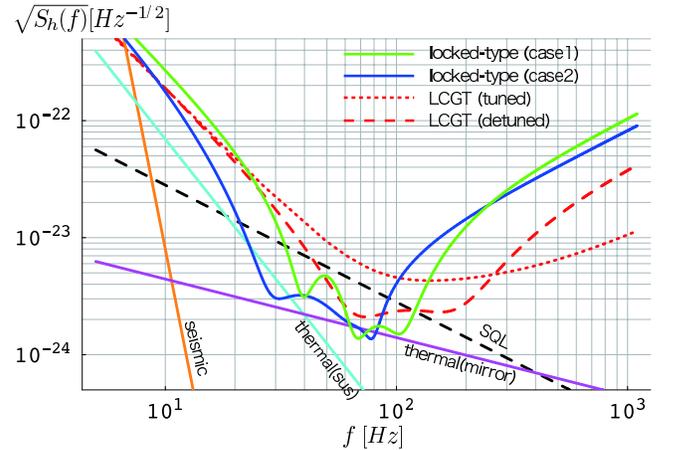}
\caption{(color online). Comparison of the sensitivity curves of a differential-type SR interferometer and LCGT. Two solid curves are the sensitivity curves of the differential-type with adjusted parameters listed in Table \ref{tab:2}. Dotted and dashed curves are the sensitivity curves of LCGT with tuned and detuned configuration, respectively.
Other straight lines are LCGT noises; thermal noise of a suspension, thermal noise of a mirror and seismic noise as indicated in the figure \cite{bib:8}. Diagonal dashed line is $h_{SQL}$.}
\label{fig:12}
\end{center}
\end{figure}

In this section, we will consider the possibility of the application of our configuration as a real GW interferometer and compare it with the next-generation ground-based interferometers such as Advanced-LIGO, which is the update version of the initial LIGO \cite{bib:10}, and LCGT, which is the next-generation ground-based interferometer in Japan and is planned to build at Kamioka mine underground \cite{bib:8}. They have much better goal sensitivity than the present detectors in operation and almost all frequency bands are limited by quantum noise. This means that there is a possibility to improve the sensitivity by reshaping and decreasing the quantum noise. In addition, since inspirals of NS-NS binary or BH-BH binary will be observed on the low frequency side, which are the main GW sources for an interferometer based on the ground, we can improve SNR significantly if we can improve the sensitivity at low frequencies by the additional third dip of a differential-type, keeping the other two dips in middle frequency. For the comparison, we should take into account not only quantum noise but also classical noise, for instance, thermal noise and seismic noise. In such a situation, the width of dips is more important than the depth of those because classical noise impairs narrow deep dips. Thus, we will compare the SNR of a differential-type for NS-NS binary or BH-BH binary with that of a recombined-type including classical noises in the sensitivity.\\
\hspace{3mm}In a differential-type SR interferometer, there are many parameters that one can adjust, FP cavity's front mirror transmissivity $T$, SR mirror reflectivity $\rho$, detuned phase in the SR cavity $\phi$, detuned phase in the dark port cavity $\theta$, and homodyne detection angle $\zeta$. For the comparison, we will fix the injected laser power to $I_0=996\;\mathrm{W}$ for the comparison with LCGT and $I_0=1284\;\mathrm{W}$ for the comparison with Advanced-LIGO. These laser powers effectively include power recycling gain and are determined to have the same laser powers in the FP cavity, $780\;\mathrm{kW}$ for LCGT and $803\;\mathrm{kW}$ for Advanced-LIGO, as in the design document \cite{bib:8, bib:10}, with the reflectivity of the FP cavity's mirror in the documents. This is because the laser power in the FP cavity is most important for quantum noise. Then, we selected the mirror transmissivity of the differential-type as $T=0.14$. In general, the larger $T$ shifts the sensitivity curve to higher frequency. This provides an advantage for a differential-type because the third additional dip can decrease noise level at low frequencies keeping the sensitivity in high frequency, though too large $T$ worsens the sensitivity. However, on the other hand, large $T$ corresponds large $I_{SQL}$ and large laser power is needed to realize good sensitivity. This is the tradeoff between $T$ and $I_{SQL}$. Therefore, we choose $T=0.14$ corresponding to $I_{SQL} \approx 2200\;\mathrm{W}$ ($I_0 \sim 0.5 I_{SQL}$ with our selection of the parameters for the comparison with both LCGT and Advanced-LIGO). We explored other parameters of a differential-type SR interferometer over all parameter space and finally selected two sets of parameters for the comparison with LCGT and one set for the comparison with Advanced-LIGO, to decrease quantum noise at low frequencies keeping the moderate sensitivity in high frequency\footnote{One can select other parameter sets to optimize the sensitivity for the particular GW sources such as NS-NS inspiral binary, BH-BH inspiral binary, binary merger and ring down, burst, stochastic background, and etc. However, for simplicity, we selected parameter sets to balance the sensitivities for all GW sources. Further detailed investigation about this is needed and is to be published.}. All parameters are listed in Table \ref{tab:2}.\\
\hspace{3mm}The sensitivity curves are shown in Fig.\ref{fig:11} and Fig.\ref{fig:12} including Advanced-LIGO and LCGT design sensitivity and other classical noise. Note that these sensitivities are calculated assuming that all optics have no loss due to absorption, scattering, etc., and that FP-cavity's end mirrors are completely reflective. However, for our purpose to compare a differential-type with a recombined-type, the assumptions are valid.\\
\hspace{3mm}SNR of a inspiral binary is given by the formula \cite{bib:7},
\begin{equation}
(SNR)^2=4\int_{0}^{\infty}df \frac{|\hat{h}(f)|^2}{S_h(f)}
\label{eqn:24}
\end{equation}
where $\hat{h}(f)$ is the Fourier component of GW amplitude and is proportional to $f^{-7/6}$ for an inspiral binary. Using this formula, one can calculate the SNR of the sensitivity curves given in Fig.\ref{fig:11} and Fig.\ref{fig:12}. However, observed frequency band for an inspiral binary is limited since it will begin to merge at the frequency corresponding to an innermost stable circular orbit. This merging frequency is given by $f_{merge}=0.02 \; c^3/(GM)$, where $G$ is the gravitational constant and $M$ is the mass of a binary star \cite{bib:7}. Taking into account this constraint, we calculated SNRs for three cases; (i) $1.4M_\odot$-$1.4M_\odot$ NS binary (full integration range of frequency), (ii) $50M_\odot$-$50M_\odot$ BH binary (limited integration range of frequency $f < 80\;\mathrm{Hz}$), and (iii) $100M_\odot$-$100M_\odot$ BH binary (limited integration range of frequency $f < 40\;\mathrm{Hz}$). The results are summarized in Table \ref{tab:3}. The values in Table \ref{tab:3} are defined as the ratio of SNR compared with Advanced-LIGO and LCGT (tuned), respectively.\\
\begin{table*}[t]
\caption{\label{tab:2}List of parameters for LCGT, Advanced-LIGO and a differential-type. We fixed the laser power $I_0=996\;\mathrm{W}$ for the comparison with LCGT and $I_0=1284\;\mathrm{W}$ for the comparison with Advanced-LIGO. These laser powers effectively include power recycling gain and are selected to have laser powers in the FP cavity, $780\;\mathrm{kW}$ for LCGT and $803\;\mathrm{kW}$ for Advanced-LIGO, given in the design document \cite{bib:8, bib:10}.}
\begin{ruledtabular}
\begin{tabular}{lccccc}
configuration & $T$ & $\rho$ & $\phi$ & $\theta$ & $\zeta$ \\
\hline 
Advanced-LIGO & 0.0707 & 0.96 & 1.51 & --- & $\pi/2$\\
Differential-type & 0.1400 &0.78 & 1.09 &1.32 & $\pi/2$\\
\hline \hline 
LCGT (tuned) & 0.0632 & 0.88 & $\pi/2$ & --- & $\pi/2$\\
LCGT (detuned) & 0.0632 & 0.95 & 1.49 &--- &0.80\\
Differential-type (case1) & 0.1400 &0.85 & 1.38 & 0.61 & 2.74\\
Differential-type (case2) & 0.1400 &0.59 &0.13 & 1.49 & 1.00\\
\end{tabular}
\end{ruledtabular}
\end{table*}
\begin{table*}[t]
\caption{\label{tab:3}The SNR sensitivity of a differential-type interferometer compared with LCGT and Advanced-LIGO. SNR for inspiral binary is calculated by the formula (\ref{eqn:24}) with integrated frequency range, all for NS binary, $f<80\;\mathrm{Hz}$ for BH binary ($50M_{\odot}$) and $f<40\;\mathrm{Hz}$ for BH binary ($100M_{\odot}$). The values in this table are defined as the ratio of SNR.}
\begin{ruledtabular}
\begin{tabular}{lccc}
configuration & NS binary & BH binary ($50M_{\odot}$) & BH binary ($100M_{\odot}$) \\
\hline 
Advanced-LIGO & 1 & 1&1\\
Differential-type & 0.90 &1.05 & 1.24\\
\hline \hline 
LCGT (tuned) & 1 & 1 &1\\
LCGT (detuned) & 1.25 & 1.56 & 1.17\\
Differential-type (case1) & 1.30 &1.87 & 1.81\\
Differential-type (case2) & 1.43 &2.28 &2.94\\
\end{tabular}
\end{ruledtabular}
\end{table*}
\hspace{3mm}In the case of Advanced-LIGO, classical noise prevents the sensitivity from improving much because the magnitude of quantum noise is comparable with that of classical noise (Fig.\ref{fig:11}). Nevertheless, SNR is improved slightly due to the third dip when the integrated frequency range is limited at low frequencies. Comparing the Advanced-LIGO, the ratio of SNR is improved by the factor of 1.24 for $100M_{\odot}$ BH binary. It could be possible to enhance the sensitivity further by using the differential-type signal recycling if the magnitude of classical noise would be decreased and, then, quantum noise would be tuned at the classical noise level. On the other hand, LCGT has the thermal noise, which is relatively smaller than that of Advanced-LIGO due to cryogenic technique. Seismic noise is also smaller because LCGT is built underground. In the case of differential-type (case2), the SNR is improved by the factor 1.43 for NS binary, 2.28 for $50M_{\odot}$ BH binary and 2.94 for $100M_{\odot}$ BH binary, compared with the SNR of LCGT (tuned). For reference, we also show the sensitivity of LCGT(detuned). Two differential-type (case1 and case2) still have better sensitivity compared with LCGT (detuned). Note that differential-type (case1) has the typical noise shape of a differential-type, and differential-type (case2) is rather special case where  three dips range. Thus, we can conclude that a differential-type has better sensitivity at low frequencies and more advantage than a recombined-type from the point of view of quantum noise .\\
\hspace{3mm}At the end of this section, we will mention the laser power needed to realize the sensitivity and the power recycling. In this calculation, we fixed the laser power $I_0=996\;\mathrm{W}$ for the comparison with LCGT and $I_0=1284\;\mathrm{W}$ for the comparison with Advanced-LIGO. For recombined-type, these laser powers are obtainable using power recycling. For a differential-type, however, these laser power is slightly large because power recycling is possible in our differential SR configuration due to SR mirrors located in front of the photo detectors, however, SR mirror's reflectivity limits the recycling gain. In other words, the large fraction of carrier light is reflected at the SR mirror and returns to the BS, however, some power is lost at the SR mirror. Therefore, several handreds watt laser is needed to achieve the laser power used in this paper if we use the SR mirror reflectivity in Table \ref{tab:2}. The power-recycling gain is not problematic if one uses a high reflective SR mirror, though it affects the sensitivity. Further detailed investigation should be done on this matter.\\

\section{conclusion and future work}
Our purpose in this paper is investigating the advanced designs of GW detectors.
We extended signal recycling scheme in a recombined-type interferometer to a differential-type. We considered the signal-recycling configuration with two SR mirrors and one CR mirror and derived the input-output relation and the spectral density. In this detector design, effective light path in the interferometer doubles compared with that of the recombined-type. As a result, this enables more dips to appear in the sensitivity curve. There are two dips due to optical resonance and one dip due to mechanical resonance.\\
\hspace{3mm}Taking advantage of this new dip, we have compared the sensitivity of a differential-type SR interferometer with a recombined-type one. We adjusted parameters of a differential-type SR interferometer and tuned it so that quantum noise is lower than classical noise. Then, we calculated SNRs for NS binary (full integrated range of frequency), $50M_{\odot}$ BH binary (limited below $80\;\mathrm{Hz}$) and $100M_{\odot}$ BH binary (limited below $40\;\mathrm{Hz}$), and compared with Advanced-LIGO and LCGT. We found that the SNRs for inspiral binaries are improved by a factor of $\approx1.43$ for NS binary, $\approx 2.28$ for $50M_{\odot}$ BH binary and $\approx 2.94$ for $100M_{\odot}$ BH binary in the case of LCGT. In the case of Advanced-LIGO, the inspiral range for NS-NS is slightly worse, however, is improved by a factor of 1.25 for $100M_{\odot}$ BH binary. As one can see from the LCGT case, there is the possibility to improve the sensitivity significantly, even if one could decrease the classical noise level by advanced techniques. Therefore, a differential-type interferometer has more advantage than recombined-type and could become a candidate for the third-generation GW interferometer.\\
\hspace{3mm}In the theoretical consideration in this paper, the differential-type SR interferometer has better sensitivity than a recombined-type one. However, we should also investigate practical aspects. Actually, what should be considered is (i) lock acquisition scheme to operate a differential-type SR interferometer, (ii) loss effects of all optics, (iii) instability of a system and ways of dealing with it. Concerning (i), the lock acquisition scheme seems complicated because there are seven mirrors that have to be controlled. We believe that, in principle, it could be operated. Concerning (ii), in this paper, we ignored all losses of optics. However, practical interferometers have losses that would affect the resonance of sideband fields and might break the dips. The effect of loss on the sensitivity should be evaluated. Concerning (iii), in general, signal-recycling systems have instabilities. This is true in the recombined-type, however, it is weak instability and can be overcome by introducing a feedback system \cite{bib:3}. The degree of instability in a differential-type interferometer has to be considered properly. Answering these questions is future work for the implementation of a differential-type SR interferometer as a real detector.

\begin{acknowledgments}
We wish to acknowledge Y.Chen, M.Ando, and S.Sato for useful discussion and comments, and R.Adhikari and T.Corbitt for useful imformation about Advanced LIGO's parameters. We also acknowledge the support of the United States National Science Foundation for the construction and operation of the LIGO Laboratory. 
\end{acknowledgments}
\appendix
\section{derivation of input-output relation in differential-type interferometer}
In this Appendix, we will derive input-output relations in a differential-type interferometer. All these formalisms are based on KLMTV's formalism \cite{bib:1}, but the notation is partly different.\\ 

\subsection{Input-output relation of conventional differential-type interferometer} 
\begin{figure*}[t]
\begin{center}
\includegraphics[width=11cm]{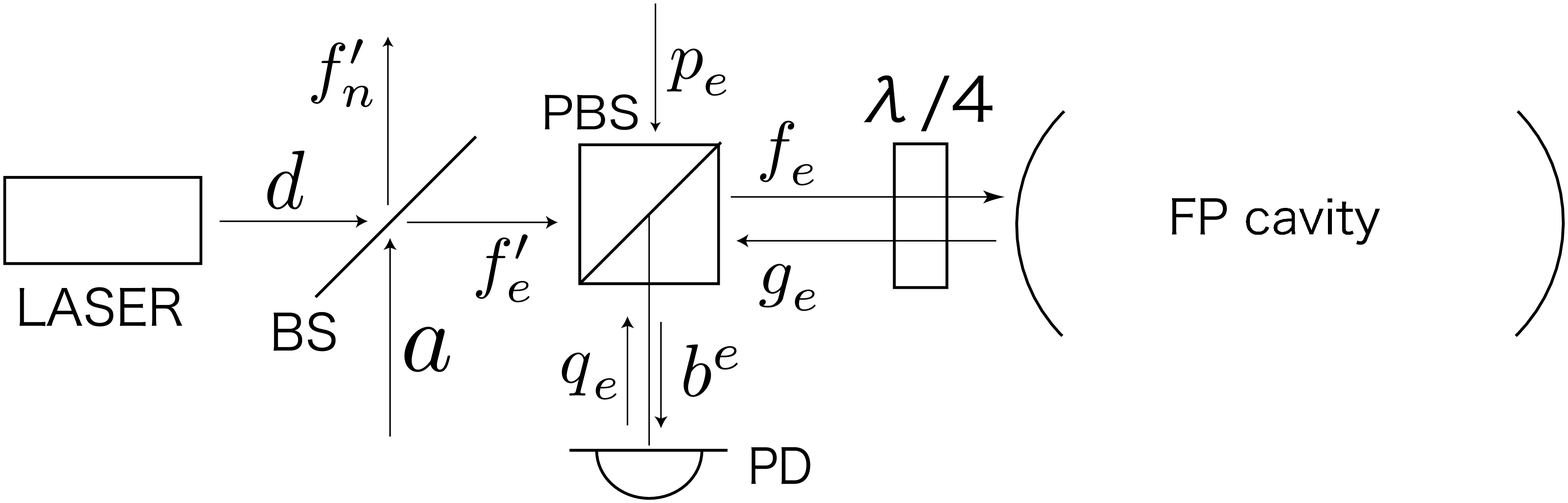}
\caption{Sideband fields in conventional differential-type interferometer.}
\label{fig:21}
\end{center}
\end{figure*}
We first consider a conventional differential-type interferometer like that in Fig.\ref{fig:1}. FP-cavity arm length is $L$. For simplicity, the lengths of other parts of the interferometer, for example, PBS-BS, PBS-photo detector and PBS-FP-cavity's front mirror, are integer times of wavelength of carrier light. Thus, there is no phase shift for carrier light in these parts. Laser power $I_0$ impinging on the beam splitter is related to the classical amplitude of the electric field $D$ by 
\begin{equation}
I_0 =  \hbar\omega_o D^2\;.
\end{equation}
$\hbar$ is the Planck constant and $\omega_0$ is the angular frequency of a carrier light. We assume, for simplicity, that all mirrors, beam splitter, PBSs have no loss due to absorption, scattering, etc. End mirrors of the FP cavities are completely reflecting mirrors. The amplitude reflectivity and transmissivity of front mirrors of the FP cavity are defined by $\{ +R,\;+T \}$ for light incident from inside the cavity, $\{ -R,\;+T \}$ for light incident from outside the cavity. $T$ and $R$ satisfy the relation $T^2+R^2=1$ \cite{bib:14}.\\
\hspace{3mm}All sideband fields including vacuum fluctuations are described in Fig.\ref{fig:21}. We shall distinguish the fields in each arm by fixing subscripts "n" and "e", which mean "north" and "east" respectively. We do not fix any "n" and "e" subscripts in general formulae, which is valid for both arms. In a conventional differential-type interferometer, there are four vacuum fields coming into the interferometer, $\mathbf{a}$, $\mathbf{d}$, $\mathbf{p}$, $\mathbf{q}$. Sideband fields $\mathbf{p}$ do not contribute to noise because the reflected $\mathbf{p}$ field at the PBS goes away from the interferometer without coupling with carrier light and the transmitted $\mathbf{p}$ field at the PBS has the polarization different from our interest output field. $\mathbf{q}$ field does not also have to be considered because the field with horizontal polarization transmits the PBS and the field with vertical polarization has different polarization from carrier light in the FP cavity and does not appear again at the photo detector. Thus, only vacuum fields we have to consider are $\mathbf{d}$ and $\mathbf{a}$.\\
\hspace{3mm}Let us express a sideband field, for instance, $\mathbf{d}$ at time $t$ using a two-photon mode \cite{bib:5,bib:6} as
\begin{eqnarray}
&&E(d;t) = \sqrt{4\pi\hbar\omega_o\over{\cal A}c} 
\nonumber\\
&&\quad\times\left\{ 
\cos(\omega_o t)
\int_0^\infty \left(d_1 e^{-i\Omega t} + 
d_1^{\dag} e^{+i\Omega t}
\right) {d\Omega\over2\pi}\right. \nonumber\\
&&\quad\quad + \left. \mathrm{\sin}(\omega_o t)
\int_0^\infty \left(d_2 e^{-i\Omega t} + 
d_2^{\dag} e^{+i\Omega t}
\right) {d\Omega\over2\pi} \right\}\;.
\label{eqn:18}
\end{eqnarray}
$d_1$, $d_2$ are field amplitudes for quadrature modes 1 and 2 and represent the amplitude of sideband. The definition is given by (\ref{eqn:6}).\\ 
\hspace{3mm}Hereafter, we will relate each sideband field at each point in the interferometer, combine them and derive the input-output relations, that is, $\mathbf{a}$-$\mathbf{b}$ relation. For convenience, we will deal with both quadrature modes with vector representation, for example, as
\begin{equation}
\mathbf{d} \equiv
\left(
\begin{array}{c}
\displaystyle
d_1 \\
\displaystyle
d_2
\end{array}
\right)\;.
\end{equation}
\begin{itemize}  
\item{at BS}\\
BS has 50:50 reflectivity, then
\begin{equation}
\mathbf{f}^{\prime}=\frac{1}{\sqrt{2}}(\mathbf{d}+\eta_{ne}\mathbf{a})\;,
\end{equation}
where $\eta_{ne}$ is $+1$ for "north" and $-1$ for "east".\\
\item{at PBS}\\
PBS reflects the light with vertical polatization and trasmits the light with horizontal polarization. Then, the relation at the PBS are
\begin{equation}
\mathbf{f}=\mathbf{f}^{\prime},\;\;\;\;\mathbf{b}=\mathbf{g}\;.
\end{equation}
\item{in FP cavity}\\ 
\hspace{3mm}Input-output relations for the FP cavity including the mirror motion due to gravitational wave and radiation pressure, are derived by KLMTV \cite{bib:1}, and we shall use the result. In their paper, no explicit expression for the following relation is given, however, combining (B4), (B21) and (B27) in Appendix in the paper, one can easily derive it and it becomes,
\begin{equation}
\mathbf{g}=e^{2i\beta}\mathbf{f}+\sqrt{K}e^{i\beta}\left[ \frac{\eta_{ne}h+(x_{BA}/L)}{h_{SQL}} \right] \mathbf{e}_d
\label{eqn:19}
\end{equation}
where 
\begin{eqnarray}
x_{BA}&=&-\frac{8\sqrt{I_0\hbar \omega_0}}{m\Omega^2 L (\gamma - i\Omega ) }(\mathbf{e}_u^T\cdot \mathbf{f}) \nonumber \\
&=&-\sqrt{K}e^{i\beta}h_{SQL}L(\mathbf{e}_u^T\cdot \mathbf{f})\;.
\label{eqn:205}
\end{eqnarray}
$m$ is the mass of FP-cavity's mirror, $h$ is the amplitude of GWs and $\beta$, $h_{SQL}$ and $K$ are defined at (\ref{eqn:34})-(\ref{eqn:36}). Here we also defined 
\begin{equation}
\mathbf{e}_u \equiv
\left(
\begin{array}{c}
\displaystyle
1 \\
\displaystyle
0
\end{array}
\right),\;\;\;\;
\mathbf{e}_d \equiv
\left(
\begin{array}{c}
\displaystyle
0 \\
\displaystyle
1
\end{array}
\right)\;.
\end{equation}
Subscript $T$ denotes the transposed matrix.\\
\end{itemize}
\hspace{3mm}Combining these equations, we can obtain the input-output relation for the conventional differential-type interferometer,
\begin{eqnarray}
b_1&=&\frac{1}{\sqrt{2}}e^{2i\beta} (d_1+\eta_{ne}a_1) \nonumber \\
b_2&=&\frac{1}{\sqrt{2}}e^{2i\beta}\left[ (d_2+\eta_{ne}a_2)-K(d_1+\eta_{ne}a_1) \right] \nonumber \\
&&+\sqrt{K}\eta_{ne}e^{i\beta}\left( \frac{h}{h_{SQL}} \right) \;.
\end{eqnarray}
Then, differentiating between "north" and "east" signals, we acquire the final expression, 
\begin{eqnarray}
\Delta b_1 &\equiv & b_1^n -b_1^e \nonumber \\
&=& \sqrt{2} e^{2 i \beta} a_1 \\
\Delta b_2 &\equiv & b_2^n -b_2^e \nonumber \\
&=& \sqrt{2} e^{2 i \beta} ( a_2 -K a_1 )+2 \sqrt{K} e^{i\beta} \left( \frac{h}{h_{SQL}}\right) \;. \nonumber \\
&&
\end{eqnarray} 

\subsection{Input-output relation of SR differential-type interferometer}
\begin{figure*}[t]
\begin{center}
\includegraphics[width=11cm]{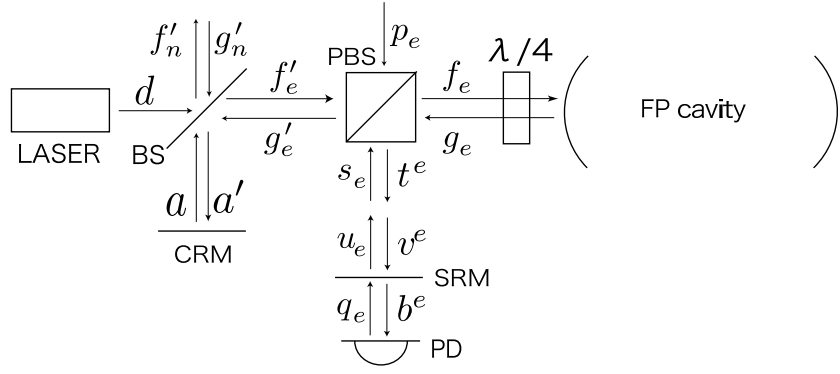}
\caption{Sideband fields in SR differential-type interferometer.}
\label{fig:22}
\end{center}
\end{figure*}
Next, we will consider the configuration of a differential-type SR interferometer like Fig.\ref{fig:2}. The assumptions about the length between each optical device and the loss of mirrors are the same as the previous subsection. SR mirrors are located in front of both the photo detectors in each arm. Amplitude reflectivity and transmissivity of SR mirror is defined by $\{+\rho,\;+\tau\}$ for the light incident from the PBS-side, $\{-\rho,\;+\tau\}$ for the light incident from the photo detector-side. They satisfy a relation $\rho^2+\tau^2 =1$. To cancel out a common mode of signals, the position of the SR mirrors is adjusted to be the same for each arm. Moreover, we also need one additional mirror, a completely reflecting mirror (CR mirror) to close the system. The position of SR and CR mirrors is characterized by parameters $\phi$ and $\theta$, which are defined by,
\begin{eqnarray}
\phi \equiv  \left[ \frac{\omega_0 \ell_s}{c}\right] _{\mod\;2\pi},\;\;\;\;\;\;\;\;\theta \equiv \left[ \frac{\omega_0 \ell_d}{c} \right] _{\mod\;2\pi}\;. \\
\nonumber  
\end{eqnarray}
where $\ell_s$ is the length between PBS and SR mirror and $\ell_d$ is the length between BS and CR mirror. We assume these lengths are small compared with the FP cavity's arm length $L$ ($\ell_s,\; \ell_d\;\sim$ several meters ). So the phase shifts for sidebands in these are negligible and we will ignore them hereafter. \\
\hspace{3mm}All sideband fields are shown in Fig.\ref{fig:22}. There are three vacuum field coming from outside, $\mathbf{d}$, $\mathbf{p}$ and $\mathbf{q}$. In this case, $\mathbf{a}$ is no longer original vacuum field because of complete reflectivity of the CR mirror. $\mathbf{p}$ does not contribute to the noise by the same reason as that of the case of a conventional differential interferometer. Thus, the input fields we need to consider are only $\mathbf{d}$ and $\mathbf{q}$.\\
\hspace{3mm}In this configuration, sideband fields and their couplings with carrier field become more complicated than a conventional one because SR and CR mirrors introduce fields with another polarization mode in the interferometer. There are two processes we need to consider. The first is that vacuum fluctuation at the dark port is no longer original vacuum fluctuation. We have to replace vacuum field $\mathbf{a}$ with the sideband field $\mathbf{a}^{\prime}$ which goes out from the beam splitter and is reflected by CR mirror. The second is that light fields with another polarization mode in the FP cavity also cause radiation pressure acting on the mirrors, since carrier light is reflected at the SR mirror and goes back to the FP cavity again. \\
\hspace{3mm}We will write down the relations of the sideband fields at each optic, combine them, and then, obtain the input-output relation.\\
\begin{itemize}
\item{at dark port}\\
\hspace{3mm}Using the expression described in (\ref{eqn:18}), the fields at BS before and after reflection by the CR mirror are related by the following equation.
\begin{equation}
E\left( a;t\right) =E\left( a^{\prime};t-2\frac{\ell_d}{c} \right)
\end{equation}   
From this equation, we can obtain,
\begin{equation}
\mathbf{a}=\mathbf{A}_d^2\mathbf{a}^{\prime}
\end{equation}
where
\begin{equation}
\mathbf{A}_d \equiv
\left(
\begin{array}{cc}
\displaystyle
\mathrm{cos}\theta & -\mathrm{sin}\theta \\
\displaystyle
\mathrm{sin}\theta & \mathrm{cos}\theta
\end{array}
\right)
\end{equation}

\item{SR mirror - PBS}\\
\hspace{3mm}Relations of sideband fields between SR mirror and PBS are given as well as the relation at the darkport by,
\begin{eqnarray}
\mathbf{v}&=&\mathbf{A}_s\mathbf{t} \\
\mathbf{u}&=&\mathbf{A}_s^{-1}\mathbf{s}
\end{eqnarray}
where
\begin{equation}
\mathbf{A}_s \equiv
\left(
\begin{array}{cc}
\displaystyle
\mathrm{cos}\phi & -\mathrm{sin}\phi \\
\displaystyle
\mathrm{sin}\phi & \mathrm{cos}\phi
\end{array}
\right)
\end{equation}

\item{at SR mirror}\\
\begin{eqnarray}
\mathbf{u}&=&\tau \mathbf{q}+\rho \mathbf{v} \\
\mathbf{b}&=&\tau \mathbf{v}-\rho \mathbf{q}
\end{eqnarray}

\item{at BS}\\
Beam splitter has 50-50 amplitude reflectivity, and then,
\begin{eqnarray}
\mathbf{f}^{\prime}&=&\frac{1}{\sqrt{2}}\left( \mathbf{d}+\eta_{ne}\mathbf{a} \right) \\
\mathbf{a}^{\prime}&=&\frac{1}{\sqrt{2}}\left( \mathbf{g}_n^{\prime}-\mathbf{g}_e^{\prime} \right)
\end{eqnarray}

\item{at PBS}\\
\hspace{3mm}In the differential-type SR configuration, different polarization mode, in other words, right-handed and left-handed mode, exist in the FP cavity. To distinguish these polarization modes, we will attach subscripts "$H$" and "$V$" to sideband fields to represent vertical polarization and horizontal polarization respectively. "$H$" and "$V$" are defined at the point between the PBS and the $\lambda/4$ plate. It should be noted that we do not have to consider sideband field $\mathbf{p}$ because it has different polarization mode from our interest field and never appears at the FP cavity and the photo detector. Thus, the relations of sidebands at PBS are
\begin{eqnarray}
\mathbf{f}^H&=& \mathbf{f}^{\prime},\;\;\;\;\;\;\mathbf{f}^V=\mathbf{s} \nonumber \\
\mathbf{g}^H&=& \mathbf{g}^{\prime},\;\;\;\;\;\;\mathbf{g}^V=\mathbf{t}\;. 
\end{eqnarray}

\item{in FP cavity}\\
\hspace{3mm}There are two polarization modes of electric fields in the FP cavity. They couple with carrier light and create radiation pressure. We can treat these couplings independently and calculate its radiation pressure. Radiation pressure acting on mirrors is the sum of contributions of two polarization modes. The relations are the same as (\ref{eqn:19}) and (\ref{eqn:205}) except for that the displacement of the mirrors are replaced by the sum of those for two polarization mode and that the laser power $I_0$ is replaced by $\rho^2 I_0$ for vertical polarization mode. Thus, the input-output relations for the FP cavity are given by
\begin{eqnarray}
\mathbf{g}^V&=&e^{2i\beta}\mathbf{f}^H \nonumber \\
&&+\sqrt{K}e^{i\beta} \left[ \frac{\eta_{ne}h+(x_{BA}^V+x_{BA}^H)/L}{h_{SQL}} \right] \mathbf{e}_d \nonumber \\
&&
\end{eqnarray}
\begin{eqnarray}
\mathbf{g}^H&=&e^{2i\beta}\mathbf{f}^V \nonumber \\
&&+\rho \sqrt{K}e^{i\beta}\left[ \frac{\eta_{ne}h+(x_{BA}^V+x_{BA}^H)/L}{h_{SQL}} \right] \mathbf{e}_d  \;, \nonumber \\
&&  
\end{eqnarray}  
where the displacements of the mirrors due to back action of radiation pressure are
\begin{eqnarray}
x_{BA}^H&=&-\sqrt{K}e^{i\beta}h_{SQL}L(\mathbf{e}_u^T \cdot \mathbf{f}^H) \\ 
x_{BA}^V&=&-\rho \sqrt{K}e^{i\beta}h_{SQL}L(\mathbf{e}_u^T \cdot \mathbf{f}^V)\;.
\end{eqnarray}
\end{itemize} 
\hspace{3mm}Combining these relations and expressing $\mathbf{b}$ with $\mathbf{q}$, after cumbersome but straightforward calculation, we can obtain, 
\begin{widetext}
\begin{eqnarray}
\Delta \mathbf{b}&=&\frac{1}{M}\left[ e^{4i\beta} 
\left(
\begin{array}{cc} \displaystyle
C_{11} & C_{12} \\
\displaystyle
C_{21} & C_{22}
\end{array}\right)
\Delta \mathbf{q}
+2 \tau \sqrt{K}e^{i\beta}
\left(
\begin{array}{c} \displaystyle
D_1  \\
\displaystyle
D_2 
\end{array}\right)
\left( \frac{h}{h_{SQL}}\right)
\right]
\\
\nonumber \\
M&=&1+\rho^2 e^{8i\beta} \nonumber \\
&&-2\rho \;e^{4i \beta} \left[ \cos2(\theta +\phi )+\frac{K}{2} \left\{ (1+\rho ^2)\;\mathrm{sin}2(\theta +\phi )+(e^{-2i\beta}+\rho ^2 e^{2i\beta})\;\mathrm{sin}2\theta +2\rho \;\mathrm{cos}2\beta\; \mathrm{sin}2\phi \right\} \right] \\
\nonumber 
\end{eqnarray}
\begin{eqnarray}
C_{11}&=& (1+\rho^2 )\;\cos2(\theta +\phi )-2\rho \;\cos4\beta \nonumber \\
&&+ \frac{K}{2}\left[ (1+\rho ^2)^2\;\mathrm{sin}2(\theta +\phi ) -\tau ^4 \;\mathrm{sin}2\theta + 2\rho \; \mathrm{cos}2\beta \{ (1+\rho ^2)\; \mathrm{sin}2\phi +2\rho \; \mathrm{sin}2\theta \} \right] \\
\nonumber \\
C_{22}&=& (1+\rho^2 )\;\mathrm{cos}2(\theta +\phi )-2\rho \;\mathrm{cos}4\beta \nonumber \\
&&+ \frac{K}{2}\left[ (1+\rho ^2)^2\;\mathrm{sin}2(\theta +\phi ) +\tau ^4 \;\mathrm{sin}2\theta + 2\rho \; \mathrm{cos}2\beta \{ (1+\rho ^2)\; \mathrm{sin}2\phi +2\rho \; \mathrm{sin}2\theta \} \right] \\
\nonumber \\
C_{12}&=&-\tau ^2 \left[ \mathrm{sin}2(\theta +\phi )+K \; \mathrm{sin}\phi \; \{ (1+\rho ^2)\;\mathrm{sin}(2\theta +\phi )+ 2\rho \; \mathrm{cos}2\beta \; \mathrm{sin}\phi \} \right] \\
\nonumber \\
C_{21}&=&\tau ^2 \left[ \mathrm{sin}2(\theta +\phi )-K \; \mathrm{cos}\phi \; \{ (1+\rho ^2)\;\mathrm{cos}(2\theta +\phi )+ 2\rho \; \mathrm{cos}2\beta \; \mathrm{cos}\phi \} \right] \\
\nonumber \\
D_1 &=& -\left[ (1+\rho^2 e^{6i\beta})\;\mathrm{sin}\phi +2\rho \;e^{3i\beta}\;\mathrm{cos}\beta \;\mathrm{sin}(2\theta +\phi ) \right] \\
\nonumber \\
D_2 &=& -\left[ (-1+\rho^2 e^{6i\beta})\;\mathrm{cos}\phi +2i\rho \;e^{3i\beta}\;\mathrm{sin}\beta \;\mathrm{cos}(2\theta +\phi ) \right]\;.\label{eqn:38}
\end{eqnarray}
\end{widetext}   
where $\Delta b \equiv b_n-b_e$, $\Delta q \equiv q_n-q_e$.

\bibliography{Locked-FP-ref}

\begin{thebibliography}{19}
\bibitem[{bib({\natexlab{a}})}]{bib:14}
\bibinfo{note}{This is different from the definition in KLMTV's paper.}
\bibitem[{bib({\natexlab{b}})}]{bib:17}
\bibinfo{note}{Circulating beam to the directions can be also done using a
  Faraday isolator and polarized carrier light. The quantum behavior of the
  interferometer with the Faraday isolator is exactly the same as that with the
  PBS and $\lambda /4$ plate discussed in this paper.}
\end{thebibliography}

\end{document}